\newcommand{\nc}{\newcommand}
\nc{\be}{\begin{equation}}
\nc{\ee}{\end{equation}}
\nc{\bea}{\begin{eqnarray}}
\nc{\eea}{\end{eqnarray}}
\nc{\beas}{\begin{eqnarray*}}
\nc{\eeas}{\end{eqnarray*}}
\nc{\noi}{\noindent}
\nc{\sD}{\not \! \! D}
\nc{\s}[1]{\not \! \!  #1}
\nc{\non}{\nonumber}
\nc{\bb}{\bibitem}
\nc{\lf}{\left}
\nc{\ri}{\right}
\nc{\mb}[1]{\makebox[#1]{}}
\nc{\pa}{\partial}
\nc{\sA}{\not \! \! A}
\nc{\newsec}[1]{\section{#1}\mb{0.5cm}}
\nc{\ra}{\rightarrow}
\nc{\la}{\leftarrow}
\nc{\mpim}{m_{\pi^-}^2}
\nc{\mpip}{m_{\pi^+}^2}
\nc{\mpi}{m_{\pi^0}^2}
\nc{\lapp}{\hbox{$ {     \lower.40ex\hbox{$<$}
                   \atop \raise.20ex\hbox{$\sim$}
                   }     $}  }
\nc{\rapp}{\hbox{$ {     \lower.40ex\hbox{$>$}
                   \atop \raise.20ex\hbox{$\sim$}
                   }     $}  }
\nc{\pieta}{$\pi^0$-$\eta,\eta'\;$}
\nc{\ep}{$e^+e^-\ra\pi^+\pi^-\;$}
\nc{\M}{{\cal M}}
\nc{\rhoom}{$\rho^0$-$\omega\;$}
\def\hhha{\rule[-3.mm]{0.mm}{7.mm}}

\documentstyle[preprint,aps,epsfig]{revtex}
\tighten
\begin{document}
\draft
\preprint{\vbox{
                                       \hfill UK/TP 98-02  \\
                                       \null\hfill hep-ph/9806423}}

\title{How Isospin Violation Mocks ``New'' Physics: \\
      $\bf \pi^0$-$\eta,\eta'$ Mixing in $\bf B\rightarrow \pi\pi$ Decays}

\author{S. Gardner\footnote{e-mail: svg@ratina.pa.uky.edu}}
\address{Department of Physics and Astronomy, \\ University of Kentucky, 
        Lexington, KY 40506-0055 USA}

\date{October 28, 1998}
\maketitle
 
\begin{abstract}

An isospin
analysis of $B\rightarrow \pi\pi$ decays 
yields $\sin 2\alpha$, where 
$\alpha\equiv {\rm arg} [-V_{td} V_{tb}^\ast/(V_{ud}V_{ub}^\ast)]$
and $V_{ij}$ is a CKM matrix element,
without hadronic uncertainty if isospin is a perfect symmetry. 
Isospin, however, is broken not only 
by electroweak effects but also 
by the $u$ and $d$ quark mass difference. 
The latter drives
$\pi^0-\eta,\eta'$ mixing and 
converts the isospin-perfect triangle relation between the $B\ra \pi\pi$
amplitudes to a quadrilateral. 
The combined isospin-violating effects impact the extracted value of 
$\sin 2\alpha$ in a significant manner, particularly if the
latter is small.

\end{abstract}
\pacs{PACS numbers: 11.30.Er, 11.30.Hv, 12.15.Hh, 13.20.He }

\bigskip
\narrowtext

In the standard model, 
CP violation is characterized by a single phase in the 
Cabibbo-Kobayashi-Maskawa (CKM) matrix, 
rendering its elements complex. 
Although CP violation
has been known in the neutral kaon system since 1964, the absence
of definitive evidence for a non-zero $\varepsilon'$ parameter leaves
the above standard model picture unsubstantiated~\cite{DGH92}. 
Indeed, probing the
precise mechanism of CP violation 
will be the primary mission of the future B factories.
The CKM matrix of the standard model is unitary, so that 
determining whether or not 
this is empirically so results in a non-trivial test of the
standard model's veracity. 
In the CKM matrix, 
only one combination of rows
and columns results in an unitarity test in which all the terms are of the 
same approximate magnitude~\cite{wolf83}; 
this is 
the unitarity triangle~\cite{pdg96}. Empirically determining whether
its angles, termed $\alpha$, $\beta$, and $\gamma$, sum to $\pi$
and whether its angles are compatible with the measured lengths of
its sides lie at the heart of these tests of the standard model. 

In the decay of a neutral $B$ meson to a CP eigenstate $f_{CP}$, 
CP violation can be generated through $B^0$-$\overline{B}^0$ 
mixing, specifically through the interference of 
$B^0 \ra f_{CP}$ and $B^0 \ra {\overline B}^0 \ra f_{CP}$.
Thus, weak phase information can be extracted 
from the time-dependent asymmetry $A(t)$, defined as
\be
A(t) \equiv
{\Gamma(B^0(t) \ra f_{CP}) - \Gamma(\overline{B}^0(t) \ra f_{CP})
\over
\Gamma(B^0(t) \ra f_{CP}) + \Gamma(\overline{B}^0(t) \ra f_{CP})} \;,
\label{asymdef}
\ee
noting $B^0(t=0)=B^0$ and ${\overline B}^0(t=0)={\overline B}^0$. 
Indeed, were only amplitudes with a single CKM phase to contribute to 
$B^0(t=0) \ra f_{CP}$, the
weak phase information could be extracted 
directly from
$A(t)$ without hadronic ambiguity~\cite{pdg96}. Unfortunately, however,
either penguin contributions
or a plurality
of tree-level contributions 
arise to cloud the above analysis~\cite{GL90}.

  Nevertheless, the 
quantity $\sin 2\alpha$, where $\alpha$ is the usual CKM angle
$\alpha\equiv 
{\rm arg} [-V_{td} V_{tb}^\ast/(V_{ud}V_{ub}^\ast)]$~\cite{pdg96},
 can be extracted without penguin ``pollution'' 
from an isospin analysis of $B\ra \pi\pi$ decays if isospin is a 
perfect symmetry~\cite{GL90}. In 
this limit, 
the Bose symmetry of the $J=0$ $\pi\pi$ state permits amplitudes 
merely of isospin  $I=0,2$. 
This implies that the amplitude 
 $B^\pm \ra \pi^\pm \pi^0$ is purely $I=2$. Thus,
as two independent amplitudes describe the three
amplitudes  $B^+ \ra \pi^+ \pi^0$, 
$B^0 \ra \pi^+ \pi^-$, and
$B^0 \ra \pi^0 \pi^0$, they 
can be 
drawn as a triangle. 
A triangle can also be formed from the
amplitudes  $B^- \ra \pi^- \pi^0$, 
${\overline B}^0 \ra \pi^+ \pi^-$, and
$\overline{B}^0 \ra \pi^0 \pi^0$, with the $B^\pm \ra \pi^\pm \pi^0$ amplitudes
forming a common base.
The strong penguin contributions are 
of $\Delta I=1/2$ character, so that they cannot contribute to the
$I=2$ amplitude and no CP violation is possible in the
$\pi^\pm\pi^0$ final states. 
This implies that the CP violation due to the penguin contribution
in $B^0 \ra \pi^+ \pi^-$, or analogously in 
${\overline B}^0 \ra \pi^+ \pi^-$, 
 can be isolated and
removed by identifying the relative magnitude and phase of the
$I=0$ and $I=2$ amplitudes.

It is our purpose to examine the manner in which
isospin-violating effects impact the extraction of 
$\sin 2\alpha$ 
as determined in $B\ra \pi\pi$ decays~\cite{GL90}.
In the standard model, isospin is an approximate symmetry. 
Isospin is broken not only by 
electroweak effects but also by the strong interaction through
the $u$ and $d$ quark mass difference. 
Both sources of isospin violation generate $\Delta I=3/2$
penguin contributions, but the latter also drives
$\pi^0-\eta,\eta'$ mixing~\cite{leut96}, admitting an $I=1$ 
amplitude~\cite{epsprime}.
These latter 
contributions  
convert the triangle relations 
between the
amplitudes to quadrilaterals. 
The effect of electroweak penguins 
has been studied earlier in
the literature, and is estimated 
to be small~\cite{deshe95,gronau95,fleischer96}.
Nevertheless, when all the effects of isospin violation are included,
the impact on the extracted value of $\sin2\alpha$ is significant.

  To review the isospin analysis
in $B\ra \pi\pi$ decays~\cite{GL90}, let us consider 
the time-dependent asymmetry $A(t)$~\cite{pdg96}:
\be
A(t) = {( 1 - |r_{f_{CP}}|^2) \over ( 1 + | r_{f_{CP}}|^2)}
\cos(\Delta m\, t) 
- {2 ({\rm Im}\, r_{f_{CP}}) 
\over 
( 1 + | r_{f_{CP}}|^2)}
\sin (\Delta m\, t) \;, 
\ee
where $r_{f_{CP}} = ({V_{tb}^\ast V_{td} / V_{tb}V_{td}^\ast})
({{\overline A}^{f_{CP}}/ A^{f_{CP}}}) \equiv 
e^{-2i\phi_m} {{\overline A}^{f_{CP}} \over A^{f_{CP}}} $ and 
$A^{f_{CP}}\equiv A(B_d^0\ra f_{CP})$. We assume the mass eigenstates
$B_L$ and $B_H$ have the same width and 
a mass difference $\Delta m\equiv B_H - B_L$. The $\sin(\Delta m\, t)$
term, resulting from $B^0$-${\overline B}^0$ mixing,
is linear in $r_{f_{CP}}$ and thus is of especial interest.  
If $f_{CP}=\pi\pi$, then the presence of penguin contributions implies 
$A^{f_{CP}}\ne {\overline A}^{f_{CP}}$. 
We denote the
amplitudes  $B^+ \ra \pi^+ \pi^0$, 
$B^0 \ra \pi^0 \pi^0$, and $B^0 \ra \pi^+ \pi^-$, 
by $A^{+0}$, $A^{00}$, and $A^{+-}$, respectively, 
and, following Ref.~\cite{GL90}, we write
\be 
{1\over 2} A^{+-} = A_2 - A_0 \quad;  A^{00} = 2A_2 + A_0 \quad; 
{1\over \sqrt{2}} A^{+0} =3 A_2 \;,
\label{triangle}
\ee
noting analogous relations for 
$A^{-0}$, ${\overline A}^{00}$, and ${\overline A}^{+-}$
in terms of ${\overline A}_2$ and ${\overline A}_0$. 
Thus, 
\be
r_{\pi^+\pi^-}= e^{-2i\phi_m} {({\overline A}_2 - {\overline A}_0)
\over
({A}_2 - {A}_0)} = e^{2i\alpha} {(1 - {\overline z})\over (1 - z)}\;,
\label{rdef}
\ee
where $z ({\overline z}) \equiv A_0/A_2 ({\overline A}_0/{\overline A}_2)$
and ${\overline A}_2/A_2 = \exp(-2i \phi_t)$ with
$\phi_t \equiv {\rm arg} ( V_{ud} V_{ub}^\ast)$ and 
$\phi_m + \phi_t = \beta + \gamma = \pi - \alpha$ in the 
standard model~\cite{pdg96}.  
Given 
$|A^{+-}|$, $|A^{00}|$, 
$|A^{+0}|$, and their charge conjugates,
the measurement of ${\rm Im}\, r_{\pi^+\pi^-}$ determines $\sin 2\alpha$,
modulo discrete ambiguities in ${\rm arg} ((1-{\overline z})/(1-z))$.
The latter can be removed via a measurement of 
${\rm Im}\, r_{\pi^0\pi^0}$ as well~\cite{GL90}. 

  We proceed by computing the individual amplitudes using the 
$\Delta B=1$ effective Hamiltonian resulting from the operator
product expansion in 
QCD in next-to-leading logarithmic (NLL) order~\cite{ali98,deshe95}, using 
the factorization approximation for the hadronic matrix elements.
The factorization approximation, which
assumes the four-quark-operator matrix elements to be saturated by vacuum 
intermediate states, finds theoretical justification in the
large $N_c$ limit of QCD~\cite{buras86} 
and phenomenological justification in comparison
with empirical branching ratios~\cite{neubert97}; 
nevertheless, it is heuristic. 
We adopt it in order to construct concrete estimates of the
effects of isospin violation in the decays of interest. 
In this context, 
we can then apply the isospin analysis delineated above to
infer $\sin 2\alpha$ and thus estimate its theoretical systematic error, 
incurred through the neglect of isospin violating
effects. 

The effective Hamiltonian ${\cal H}^{\rm eff}$ 
for $b\ra d q\overline q$ 
can be parametrized as~\cite{ali98}
\be
{\cal H}^{\rm eff}= 
{G_F \over \sqrt{2}}\left[ V_{ub}V^*_{ud} (C_1 O_1^u + C_2 O_2^u)
+ V_{cb}V^*_{cd} (C_1 O_1^c + C_2 O_2^c)
 - V_{tb}V^*_{td} \left(\sum_{i=3}^{10} C_i O_i + C_g O_g\right) 
\right] \;,
\ee
where $O_i$ and $O_g$ are as per Ref.~\cite{ali98}; 
we also adopt their Wilson
coefficients $C_i$ and $C_g$, computed in the naive
dimensional regularization scheme at a renormalization scale 
of $\mu=2.5$ GeV~\cite{ali98}. 
In NLL order, the Wilson coefficients are scheme-dependent;
yet, after computing the hadronic matrix elements to one-loop-order,
the matrix elements of the effective Hamiltonian are  still
scheme-independent~\cite{fleis93}. This 
can be explicitly realized
through the replacement 
$\langle dq{\overline q}| {\cal H}^{\rm eff} | b\rangle 
= (G_F/\sqrt{2}) \langle dq{\overline q} | 
[ V_{ub}V^*_{ud} (C_1^{\rm eff} O_1^u + C_2^{\rm eff} O_2^u)
 - V_{tb}V^*_{td} \sum_{i=3}^{10} C_i^{\rm eff} O_i ] | b \rangle^{\rm tree}$,
where ``tree'' denotes a tree-level matrix element and the 
$C_i^{\rm eff}$ are from Ref.~\cite{ali98}. 
The $C_i^{\rm eff}$ are complex~\cite{bss79} and 
depend on both the CKM matrix 
parameters
and $k^2$, where $k$ is the momentum transferred to the
$q\overline q$ pair in $b\ra d q\overline q$. Noting Ref.~\cite{wolf83}
we use $\rho=0.12$, $\eta=0.34$, and $\lambda=0.2205$~\cite{ali98,SMparam}
unless otherwise stated. 
One expects $m_b^2/4 \lapp k^2 \lapp m_b^2/2$~\cite{kinem}; we 
use $k^2/m_b^2 = 0.3, 0.5$ in  what follows. 

To include the effects of \pieta mixing, we write the
pion mass eigenstate $|\pi^0\rangle$ 
in terms of the $SU(3)_f$ perfect states
$|\phi_3\rangle=|u{\overline u} - d{\overline d}\rangle/\sqrt{2}$, 
$|\phi_8\rangle=|u{\overline u} +
 d{\overline d} - 2s{\overline s}\rangle/\sqrt{6}$, 
and 
$|\phi_0\rangle=
|u{\overline u} + d{\overline d} + s{\overline s}\rangle/\sqrt{3}$.
To leading order in isospin violation~\cite{leut96} 
\be
|\pi^0 \rangle = |\phi_3\rangle + \varepsilon 
(\cos \theta |\phi_8\rangle - \sin \theta |\phi_0\rangle)
+ \varepsilon' (\sin \theta |\phi_8\rangle + \cos \theta
|\phi_0\rangle) \;,
\label{physpi}
\ee
where $|\eta\rangle=\cos \theta |\phi_8\rangle - \sin \theta |\phi_0\rangle
+ O(\varepsilon)$, and $|\eta'\rangle=\sin \theta |\phi_8\rangle + 
\cos \theta |\phi_0\rangle + O(\varepsilon')$. 
Expanding QCD to leading order in $1/N_c$, 
momenta, and quark masses permits 
the construction of a low-energy, effective Lagrangian in which 
the pseudoscalar meson octet and singlet states are 
treated on the same footing~\cite{leut96,DGH92}. Diagonalizing its
quadratic terms in $\phi_3$, $\phi_8$, and $\phi_0$ determines 
the mass eigenstates $\pi^0$, $\eta$,
and $\eta'$ and yields
$\varepsilon = \varepsilon_0 \chi \cos \theta$ and 
$\varepsilon' = -2\varepsilon_0 \tilde\chi  \sin \theta$,
where $\chi= 1 + (4m_K^2 - 3m_{\eta}^2 - m_{\pi}^2)/(m_\eta^2 - m_\pi^2)
\approx 1.23$, $\tilde\chi = 1/\chi$, 
$\varepsilon_0 \equiv \sqrt{3}(m_d - m_u) /(4(m_s - \hat{m}))$,
 and $\hat{m}\equiv(m_u + m_d)/2$~\cite{leut96}. Thus the magnitude
of isospin breaking is controlled by the SU(3)-breaking parameter 
$m_s - \hat{m}$. The $\eta$-$\eta'$ mixing angle $\theta$ 
is found to be
$\sin 2\theta= - (4\sqrt{2}/3)(m_K^2 - m_\pi^2)/(m_{\eta'}^2 - m_\eta^2)
\approx -22^{\circ}$~\cite{leut96}. 
The resulting $\varepsilon
=1.14\varepsilon_0$ is comparable
to the one-loop-order chiral perturbation theory 
result of $\varepsilon=1.23\varepsilon_0$
in $\eta\ra \pi^+ \pi^- \pi^0$~\cite{gasser85,leut96}.
Using $m_q(\mu=2.5\, {\rm GeV})$ from Ref.~\cite{ali98}, we find
$\varepsilon = 1.4 \cdot 10^{-2}$ and
$\varepsilon' = 7.7 \cdot 10^{-3}$. 

\begin{figure}
\centerline{\epsfig{file=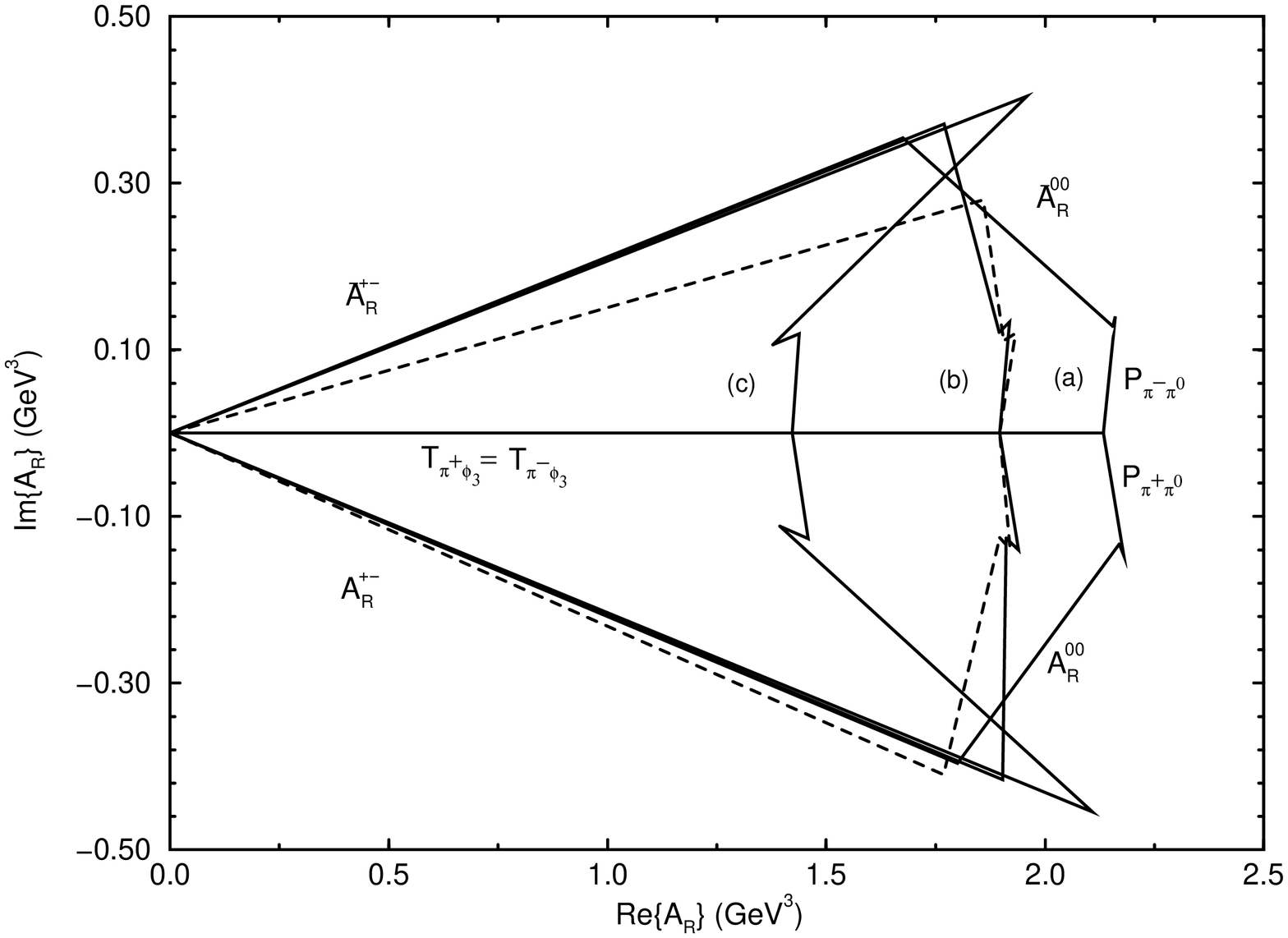,angle=-90,width=5.0in}}
\vspace{100pt} 
\caption{
Reduced amplitudes 
in $B\ra\pi\pi$ 
in the factorization 
approximation with 
[$N_c$, $k^2/m_b^2$] for  
a) [2,0.5], b) [3, 0.5] (solid line) and [3, 0.3] (dashed line), and
c) [$\infty$,0.5]. 
}
 \label{figone}
\end{figure}

We now compute 
the matrix elements of the above effective
Hamiltonian in the cases of interest. 
We define the decay constants
$\langle \pi^- (p)| {\overline d} \gamma_\mu \gamma_5 u | 0 \rangle 
\equiv -i f_{\pi} p_{\mu}$ and 
$\langle \phi_i (p)| {\overline u} \gamma_\mu \gamma_5 u | 0 \rangle 
\equiv -i f_{\phi_I}^u p_{\mu}$, 
and use SU(3)$_f$ to relate $f_{\phi_i}^q$ to $f_{\pi}$.
Finally, using 
the quark equations of motion with PCAC and introducing
$a_i\equiv C_i^{\rm eff} + C_{i+1}^{\rm eff}/N_c$ for $i$ odd
and $a_i\equiv C_i^{\rm eff} + C_{i-1}^{\rm eff}/N_c$ for $i$ even,
the $B^-\ra \pi^-\phi_3$ 
matrix element in the factorization approximation with use of
the Fierz relations~\cite{DGH92} is 
\bea
&\langle& \pi^- \phi_3 | 
{\cal H}^{\rm eff} | B^-
\rangle 
= {G_F\over \sqrt{2}}
[ V_{ub}V^*_{ud} (i f_{\pi} F_{B^-\phi_3}(m_{\pi^-}^2)a_1
+ i f_{\phi_3}^u F_{B^-\pi}(m_{\pi^0}^2)a_2) 
- V_{tb}V^*_{td} \nonumber \\
&\times& 
(i f_{\pi} F_{B^-\phi_3}(m_{\pi^-}^2) 
(a_4 + a_{10} + {2m_{\pi^-}^2(a_6 + a_8)\over (m_u+m_d)(m_b - m_u)})
-
i f_{\phi_3}^u F_{B^-\pi}(m_{\pi^0}^2) \\
&\times& 
(a_4 + {3\over 2}(a_7 - a_9) - {1\over 2}a_{10} 
+ {m_{\pi^-}^2(a_6 - {1\over 2}a_8)\over m_d(m_b - m_d)})
)]\;. \nonumber 
\eea
The 
transition form factors are given by
$F_{B^-\pi}(q^2)= 
(m_{B^-}^2 - m_{\pi^-}^2) F_0^{B\ra \pi}(0)/(1 - q^2/M_{0^+}^2)$,
where we use $F_0^{B\ra \pi}(0)=0.33$ and 
$M_{0^+}=5.73$ GeV as per Refs.~\cite{ali98,bsw87}. Also
$F_{B\phi_3}= F_{B\pi}/\sqrt{2}$,
$F_{B\phi_8}= F_{B\pi}/\sqrt{6}$, and
$F_{B\phi_0}= F_{B\pi}/\sqrt{3}$.
In the presence of isospin violation, 
the $B^-\ra \pi^- \phi_3$ amplitude is no longer purely
$I=2$. However, 
if $\pi^0$-$\eta,\eta'$ mixing is neglected,
${\overline A}^{\,+-} + 2{\overline A}^{\,00} = 
\sqrt{2}\,{\overline A}^{\,-0}$,
from Eq.~\ref{triangle}, 
is still satisfied, as we ignore 
the small mass
differences $m_{\pi^{\pm}}-m_{\pi^0}$ and $m_{B^{\pm}}-m_{B^0}$. 
Now with $\pi^0$-$\eta,\eta'$ mixing, 
\begin{mathletters}
\bea
 A^{-0}&=& \langle \pi^- \phi_3 | {\cal H}^{\rm eff} | B^-\rangle 
+ \varepsilon_8\langle \pi^- \phi_{8} | 
{\cal H}^{\rm eff} | B^-\rangle 
+ \varepsilon_0\langle \pi^- \phi_{0} | 
{\cal H}^{\rm eff} | B^-\rangle \\
{\overline A}^{\,00}&=& \langle \phi_3 \phi_3 | {\cal H}^{\rm eff} | 
{\overline B}^0 \rangle 
+ 2\varepsilon_8\langle \phi_3 \phi_{8} | 
{\cal H}^{\rm eff} | {\overline B}^0 \rangle 
+ 2\varepsilon_0\langle \phi_3 \phi_{0} | 
{\cal H}^{\rm eff} | {\overline B}^0 \rangle \;,
\eea
\label{amps}
\end{mathletters}
where $\varepsilon_8\equiv \varepsilon \cos\theta + \varepsilon' \sin\theta$,
$\varepsilon_0\equiv \varepsilon' \cos\theta - \varepsilon \sin\theta$, and 
further details appear in Ref.~\cite{svg98}.
The $B\ra \pi\pi$ amplitudes now satisfy
\bea
{\overline A}^{\,+-} + 2{\overline A}^{\,00} 
&-& \sqrt{2}\,A^{\,-0} 
= 
4\varepsilon_8 \langle \phi_3 \phi_{8} | 
{\cal H}^{\rm eff} | {\overline B}^0\rangle 
+4\varepsilon_0 \langle \phi_3 \phi_{0} | 
{\cal H}^{\rm eff} | {\overline B}^0\rangle \nonumber \\
&-& \sqrt{2} \varepsilon_8
\langle \pi^- \phi_{8} | 
{\cal H}^{\rm eff} | B^-\rangle 
- \sqrt{2} \varepsilon_0 
\langle \pi^- \phi_{0} | 
{\cal H}^{\rm eff} | B^-\rangle \;,
\label{newrel}
\eea
and thus the previous
triangle relation becomes a quadrilateral.
Numerical results in the factorization approximation
for the reduced amplitudes $A_R$ and ${\overline A}_R$, where
${\overline A}_R^{\,00} \equiv 
2{\overline A}^{\,00} /((G_F/\sqrt{2}) i V_{ub} V_{ud}^*)$,
${\overline A}_R^{\,+-} \equiv 
{\overline A}^{\,+-} /((G_F/\sqrt{2}) i V_{ub} V_{ud}^\ast)$, and
$A_R^{-0} \equiv \sqrt{2}A^{-0} /((G_F/\sqrt{2}) i V_{ub} V_{ud}^\ast)$,
with $N_c=2,3,\infty$ and $k^2/m_b^2=0.3,0.5$ 
are shown in Fig.~\ref{figone}. Here the parameter $N_c$ 
defines different hadronic models; 
$N_c=2,3$ bound the value favored 
from fits to measured branching ratios~\cite{neubert97}.
$A_R^{+0}$ and $A_R^{-0}$ are
broken into tree and penguin contributions, so that 
$A_R^{+0}\equiv T_{\pi^+\phi_3} + P_{\pi^+\pi^0}$ and 
$A_R^{-0}\equiv T_{\pi^-\phi_3} + P_{\pi^-\pi^0}$, where 
$P_{\pi^\pm\pi^0}$ is defined to include the isospin-violating tree
contribution in $A_R^{\pm0}$ as well. The shortest side in each
polygon is the vector defined by the RHS of Eq.~\ref{newrel}.
For reference, note that the
ratio of penguin to tree amplitudes in $B^-\ra \pi^-\pi^0$ is
$|P|/|T| \sim (2.2 - 2.7)\% | V_{tb}V^*_{td}| / |V_{ub}V^*_{ud}|$ for 
$N_c=2,3$ and $k^2$ above. 
Were electroweak penguins the only source of isospin violation, then
$|P|/|T| \sim (1.4-1.5)\% | V_{tb}V^*_{td}| / |V_{ub}V^*_{ud}|$,
commensurate with the estimate of 1.6\% in Ref.~\cite{deshe95}. 
Following Ref.~\cite{GL90}, 
the determination of ${\rm Im}\, r_{\pi^+\pi^-}$ 
yields $\sin 2\alpha$,
modulo 
discrete ambiguities in $z$ and ${\overline z}$ as in Eq.~\ref{rdef}. 
However, $\sin 2\alpha$ can be determined uniquely 
through a comparison with
$\sin2\alpha$ from ${\rm Im}\, r_{\pi^0\pi^0}$, as 
only one pair of the $\sin2\alpha$ extracted
from Im$r_{\pi^+\pi^-}$ and Im$r_{\pi^0\pi^0}$ 
likely match. 
As $|{\overline A}_2|/| A_2|\ne 1$ in the presence of isospin
violation, we have retained this explicit factor in the last
equality of Eq.~\ref{rdef} in order to extract $\sin 2\alpha$
as accurately as possible. 
The values of $\sin2\alpha$ extracted from 
the amplitudes in the factorization approximation with 
$N_c$ and $k^2/m_b^2=0.5$ are shown
in Table \ref{delcf} for a variety of input values of $\sin 2\alpha$
--- the results for $k^2/m_b^2=0.3$ are similar
and have been omitted.
In the presence of $\pi^0$-$\eta,\eta'$ mixing,
the ${\overline A}_R^{\,+-}$, ${\overline A}_R^{\,-0}$, and 
${\overline A}_R^{\,00}$ amplitudes obey a quadrilateral relation 
as per Eq.~\ref{newrel}, so that the amplitudes of interest need
no longer form triangles. Consequently, 
the values of $\sin2\alpha$ 
extracted from Im$r_{\pi^+\pi^-}$ and Im$r_{\pi^0\pi^0}$ can not only differ
markedly from the value of $\sin2\alpha$ input but also need not match.
The incurred error in $\sin 2\alpha$ increases as the value 
to be extracted decreases; the structure of Eq.~\ref{rdef} suggests this,
for as $\sin 2\alpha$ decreases the quantity
Im$((1 - {\overline z})/(1 -z))$ becomes more important to 
determining the extracted value. 
It is useful to constrast the impact of 
the various isospin-violating effects.
The presence of $\Delta I=3/2$ penguin
contributions, be they from $m_u\ne m_d$ or electroweak effects, shift
the extracted value of $\sin 2\alpha$ from its input value, yet the 
``matching'' of the $\sin 2\alpha$ values in
Im$r_{\pi^+\pi^-}$ 
and Im$r_{\pi^0\pi^0}$ is
unaffected. The mismatch troubles seen in Table \ref{delcf} are driven by
$\pi^0$-$\eta,\eta'$ mixing, though the latter shifts the values of
$\sin 2\alpha$ in Im$r_{\pi^+\pi^-}$ as well. 
Picking the closest
matching values of $\sin 2\alpha$ in the two final states also 
picks the solutions closest to the input value; the exceptions are noted
in Table \ref{delcf}. The matching procedure can also yield
the wrong strong phase; in case b) of Table \ref{delcf} with
$N_c=3,\infty$, the triangles of the chosen solutions 
``point'' in the same direction, 
whereas they actually point oppositely. 

If $|A^{00}|$ and $|{\overline A}^{00}|$ 
are small~\cite{GL90} the complete isospin analysis may not be
possible, so that 
we examine the utility of the bounds proposed in Ref.~\cite{GQpi97}
on the strong phase 
$2\delta_{\rm true}\equiv {\rm arg}((1 - {\overline z})/(1 - z))$
in Eq.~\ref{rdef}. The bounds 
$2\delta_{\rm GQI}$ and $2\delta_{\rm GQII}$ given by 
Eqs.~2.12 and 2.15, respectively, in Ref.~{\cite{GQpi97}} 
follow from Eq.~\ref{triangle}, and thus can be broken
in the presence of isospin violation. As shown in 
Table \ref{delcf}, the bounds typically are broken, and their 
efficacy does not improve as the value of $\sin 2\alpha$ 
to be reconstructed grows large. 
  
  To conclude, we have considered the role of isospin violation
in $B\ra\pi\pi$ decays and have found the effects to be significant. 
Most particularly, the utility of the
isospin analysis in determining $\sin 2\alpha$ strongly depends 
on the value to be reconstructed. The error in $\sin 2\alpha$
from a Im$r_{\pi^+\pi^-}$ measurement can be 50\% or more for
the small values of
$\sin 2 \alpha$ currently favored by 
phenomenology~\cite{ali98,SMparam,DLcomm}; however, 
if $\sin 2\alpha$ were
near unity, the error would decrease to less than 10\%. 
The effects found arise in part because the penguin contribution in 
$B^0 \rightarrow \pi^+ \pi^-$, e.g., is itself small; we estimate
$|P|/|T| < 9\% | V_{tb}V^*_{td}| / |V_{ub}V^*_{ud}|$. Relative to this
scale, the impact of \pieta mixing is significant. Yet, 
were the penguin
contributions in $B\rightarrow \pi\pi$ larger, the isospin-violating
effects considered would still be germane, for not only would 
the $\Delta I=3/2$ penguin contributions likely be larger but 
the $B\rightarrow \pi\eta$ and
$B\rightarrow \pi\eta'$ contributions could also be 
larger as well~\cite{ciuchini97}. To conclude, we have shown that 
the presence of $\pi^0$-$\eta,\eta'$ mixing breaks the
triangle relationship, Eq.~\ref{triangle}, usually assumed~\cite{GL90} 
and can mask the true value of $\sin2\alpha$.

S.G. thanks S. Cotanch for a suggestive query and D. London,
T. Cohen, A. Kagan, A.~W. Thomas, and M. Wise for helpful discussions. 
This work was supported by the DOE under DE-FG02-96ER40989.


\begin{table}[delcf]
\begin{center}

\caption{
Strong phases and inferred values of $\sin2\alpha$~\protect{\cite{GL90}} 
from amplitudes in the factorization
approximation with $N_c$ and $k^2/m_b^2=0.5$. 
The strong phase $2\delta_{\rm true}$
is the opening
angle between the ${\overline A_R}^{+-}$ and ${A_R}^{+-}$ amplitudes
in Fig.~1, 
whereas 2$\delta_{\rm GL}$ is the strong phase associated
with the closest matching $\sin2\alpha$ values, denoted 
$(\sin 2\alpha)_{\rm GL}$, from 
Im$r_{\pi^+\pi^-}$/Im$r_{\pi^0\pi^0}$, respectively. 
The bounds $|2\delta_{\rm GQI}|$ and $|2\delta_{\rm GQII}|$  on 
$2\delta_{\rm true}$ 
from Eqs.~2.12 and 2.15 of Ref.~\protect{\cite{GQpi97}} are also shown.
All angles are in degrees.
We input a) $\sin2\alpha=0.0432$~\protect{\cite{ali98,SMparam}},
b) $\sin2\alpha=-0.233$ ($\rho=0.2,\eta=0.35$)~\protect{\cite{DLcomm}}, and 
c) $\sin2\alpha=0.959$ ($\rho=-0.12$).
$^\ast$ The matching
procedure fails to choose a $\sin2\alpha$ which is as 
close to the input value as possible. $^\dagger$ 
The discrete ambiguity in the strong phase is resolved wrongly.
}

\begin{tabular}{ccccccc}

\hhha 
case & $N_c$ & 2$\delta_{\rm true}$ &  $|2\delta_{\rm GQI}|$ & 
$|2\delta_{\rm GQII}|$ &  $|2\delta_{\rm GL}|$ &  $(\sin 2\alpha)_{\rm GL}$  \\
\hline
a & 2 &  24.4 & 26.1 & 15.8 & 16.6 & -0.0900/-0.0221 \\
a & 3 &  24.2 & 16.9 & 16.1  & 16.2  & -0.0926/0.107 \\
a & $\infty$ & 23.8  & 59.4  & 25.1  & 23.6 & 0.0451/0.394 \\
b & 2 &  19.6 & 23.4 & 12.1 & 12.9 & -0.343/-0.251 \\
b & 3 &  19.4 & 13.5 & 12.9  & 13.0  & -0.719/-0.855($\ast$)\\
b & $\infty$ & 19.2  & 59.9  & 23.6  & 0.76 & -0.550/-0.814($\ast,\dagger$) \\
c & 2 &  28.3 & 36.5 & 20.4 & 21.0 & 0.917/0.915 \\
c & 3 &  28.0 & 24.0 & 19.1  & 19.0  & 0.905/0.952 \\
c & $\infty$ & 28.3  & 36.5  & 20.4  & 21.0 & 0.917/0.915 
\end{tabular}
\label{delcf}
\end{center}
\end{table}


\begin{references}

\bibitem{DGH92} J.F. Donoghue, E. Golowich, and B.R. Holstein,
{\em Dynamics of the Standard Model} (Cambridge University Press, Cambridge, 
1992).

\bibitem{wolf83} L.~Wolfenstein, Phys. Rev. Lett.~{\bf 51}, 1945 (1983).
We use Eq.(10) to compute the CKM matrix elements.

\bibitem{pdg96} R.M. Barnett {\it et al.}, 
Phys. Rev. D~{\bf 54}, 1 (1996).

\bibitem{GL90} M. Gronau and D. London, Phys. Rev. 
Lett.~{\bf 65}, 3381 (1990).
Note $A^{+-}_{\rm GL}=A^{+-}/\sqrt{2}$ and 
$A^{\pm 0}_{\rm GL}=A^{\pm 0}/\sqrt{2}$.

\bibitem{leut96} H. Leutwyler, Phys. Lett. {\bf B374}, 181 (1996).

\bibitem{epsprime} $\pi^0$-$\eta,\eta'$ mixing also 
affects $\epsilon'/\epsilon$, see A.J. Buras and J.-M. Gerard, 
Phys. Lett. B {\bf 192}, 156 (1987); 
J.F. Donoghue, E. Golowich, B.R.
Holstein, and J. Trampetic, Phys. Lett. B {\bf 179}, 361 (1986).

\bibitem{deshe95}  N.G. Deshpande and X.-G. He, 
Phys. Rev. Lett.~{\bf 74}, 26 (1995). For
$c_1 - c_6$ note also A.J. Buras, M. Jamin, M.E. Lautenbacher,
and P.H. Weisz, Nucl. Phys.~B~{\bf 370}, 69 (1992); {\bf 375}, 501 (1992).

\bibitem{gronau95} M. Gronau, O.F. Hern\'andez, D. London, and J.L. Rosner,
Phys. Rev. D~{\bf 52}, 6374 (1995).

\bibitem{fleischer96} R. Fleischer, Phys. Lett. B {\bf 365}, 399 (1996).

\bibitem{ali98} A. Ali, G. Kramer, and C.-D. L\"u, 
Phys. Rev. D {\bf 58}, 094009 (1998).

\bibitem{buras86} A.J. Buras, J.-M. G\'erard, and R. R\"uckl, 
Nucl. Phys.~B~{\bf 268}, 16 (1986). 

\bibitem{neubert97} M. Neubert and B. Stech, hep-ph/9705292, 
to appear in {\em Heavy Flavors, $2^{nd}$ Ed.}, edited by A.J.
Buras and M. Lindner (World Scientific, Singapore).

\bibitem{fleis93} R. Fleischer, Z. Phys.~C {\bf 58}, 483 (1993).

\bibitem{bss79} M. Bander, D. Silverman, and A. Soni, 
Phys. Rev. Lett. ~{\bf 43}, 242 (1979). 

\bibitem{SMparam} A.~Ali, hep-ph/9801270;
A. Ali and D. London, Nucl.~Phys. B (Proc. Suppl.) {\bf 54A}, 297 (1997);
F.~Parodi, P.~Roudeau, and A.~Stocchi, hep-ph/9802289;
S. Mele, hep-ph/9810333.

\bibitem{kinem} N.G. Deshpande and J. Trampetic, 
Phys. Rev. D~{\bf 41}, 2926 (1990);
H. Simma and D. Wyler, Phys. Lett. {\bf B272}, 395 (1991).

\bibitem{gasser85} J. Gasser and H. Leutwyler, Nucl. 
Phys.~B~{\bf 250}, 539 (1985). 

\bibitem{bsw87} M.~Bauer, B.~Stech, and M.~Wirbel, Z. Phys.~C {\bf 29}, 637 
(1985); {\bf 34}, 103 (1987). 

\bibitem{svg98} S. Gardner, in preparation.

\bibitem{GQpi97}  Y. Grossman and H.R. Quinn, 
Phys. Rev. D~{\bf 58}, 017504 (1998). 

\bibitem{DLcomm} D. London, private communication. The 
$\rho$ and $\eta$ given are the new central values from an update
of Ref.~{\cite{SMparam}}. 

\bibitem{ciuchini97} M. Ciuchini et al., 
Nucl. Phys. B 501 (1997), 271; 512 (1998), 3. 




\end{references}
\end{document}